\documentclass[aps,amssymb,floatfix,prd,preprintnumbers]{revtex4-2}
\usepackage{epstopdf}
\usepackage{capt-of}
\usepackage{graphicx}  
\usepackage{dcolumn}   
\usepackage{bm}
\usepackage{amsmath}
\usepackage[font=scriptsize]{caption}
\usepackage[colorlinks]{hyperref}
\begin{document}
\input epsf.tex
\title{\bf Dynamics of quasi de Sitter and linear combination of exponential models in extended gravity}

\author{B. Mishra\footnote{Department of Mathematics, Birla Institute of Technology and Science-Pilani, Hyderabad Campus, Hyderabad-500078, India, email: bivu@hyderabad.bits-pilani.ac.in}, Eesha Gadia\footnote{Department of Mathematics, Birla Institute of Technology and Science-Pilani, Hyderabad Campus, Hyderabad-500078, India, email: eesha.gadia123@gmail.com}, S.K. Tripathy \footnote{Department of Physics, Indira Gandhi Institute of Technology, Sarang, Dhenkanal, Odisha-759146, India, tripathy\_sunil@rediffmail.com}}

\affiliation{ }

\begin{abstract}
\textbf{Abstract}
In this paper, we have studied the dynamical aspects of some cosmological models in an isotropic space-time within the framework of an extended gravity theory. Two accelerating cosmological models are presented with quasi-de Sitter (QDS) and linear combination of exponential (LCE) scale factors. The geometrical testing mechanism has been performed and analysed. While the results of the QDS model show a departure from the concordant $\Lambda$CDM model, the resulting behavior of the LCE model seems to be compatible with the $\Lambda$CDM model only at the present epoch and is incompatible in the near past.
\end{abstract}

\maketitle
\textbf{Keywords}: Isotropic space-time, Modified gravity, Stability analysis, Geometrical diagnostic.

\section{Introduction}

In modern cosmology, the main theme of research is on the issue of late time cosmic acceleration phenomena. Supernovae type Ia provide strong evidence that the Universe is currently undergoing an accelerated phase of expansion. Plethora of observations during the last two decades have confirmed the late time cosmic acceleration of the Universe \cite{Perlmutter98,Reiss98,Knop03,Eisenstein05,Spergel07,Sullivan11,Suzuki12}. These
observations have developed a curiosity among the cosmologists to explain this late time dynamics. In the purview of General Relativity (GR), it becomes difficult to explain this issue. Hence the idea of modifying GR has taken momentum in the last decade. Researchers changed  both the geometrical and matter part of field equations of GR. The changes made in the matter part lead to the inclusion of some exotic form of energy known as dark energy and the changes in the geometric part lead to the extended gravity. Several geometrical extensions of GR have been made in recent past such that $f(R)$ gravity \cite{Nojiri06a,Capozziello11,Nojiri17}, $f(R,T)$ gravity \cite{Harko11}, $f(R,G)$ gravity \cite{Nojiri06b}, $f(T)$ gravity \cite{Obukhov03}, $f(Q,T)$ gravity \cite{Xu19}. The $f(R,T)=f_1(R)+f_2(T)$ gravity consists of the functions of Ricci scalar $R$ and trace of energy momentum tensor $T$ and have been suggested in three different functional forms \cite{Harko11} such as (a) $R+2f(T)$; (b) $f_1(R)+f_2(T)$ and (c) $f_1(R)+f_2(R).f_3(T)$. Each functional has its own merit to answer the late time cosmic acceleration issue. It is believed that, the dark sector of the Universe is responsible for the cosmic evolution and the dynamics at least at its late phase. The dark sector responsible for cosmic evolution may be the dark matter or the dark energy. It still remains as an open question, whether the dark matter is mediated through a weakly interacting massive particle or  is a manifestation of GR modification. The dark energy on the other hand is an exotic energy form that violates the strong energy condition and is associated with a fluid of negative pressure. Within the purview of GR, the dark energy is required to explain the late time cosmic speed up phenomena. Till date, the quest to unfold the mystery of late time cosmic acceleration through different dark energy models and modified gravity theories still remains unquenched. Besides, there have been questions and possible efforts to find solutions concerning the initial singularity problem. In this context, there have been many interesting discussions in recent times \cite{Odintsov2020, Odintsov2020a, Odintsov2021, Oikonomou2021, Oikonomou2014}.

Several cosmological models under $f(R,T)$ gravity have been proposed in last decade. We have discussed some prominent models that are successful in giving some insight to the issue of late time cosmic acceleration phenomena. Houndjo \cite{Houndjo12} has investigated the cosmological reconstruction with the power law cosmology. Santos \cite{Santos13} obtained the Godel solution and Azizi \cite{Azizi13} studied the wormhole geometry in the $f(R,T)$ gravity. In the metric formalism, Alavarenga et al. \cite{Alvarenga13} have studied the scalar perturbations. Pasqua et al. \cite{Pasqua13} have shown the quintessence behaviour of modified holographic dark energy model. Yadav \cite{Yadav14} has shown the existence of string cosmological model in power law cosmology. Considering the existence of conformal killing vector, Das et al. \cite{Das16}  have shown the behaviour of compact stars in $f(R,T)$ gravity. Yousaf et al. \cite{Yousaf16} have examined the irregularity factors of spherical star in anisotropic and isotropic fluid. Tripathy et al. \cite{Tripathy2019} have discussed some cosmological models speculating a bouncing scenario in $f(R,T)$ gravity. Zaregonbadi et al. \cite{Zaregonbadi16} derived the metric components in the galactic halo with the minimal coupling of matter and geometry. Alhamzawi and Alhamzawi \cite{Alhamzawi16} have calculated the effects of $f(R,T)$ gravity in gravitational lensing and compared with the effects of general relativity. Fayaz \cite{Fayaz16} have analysed the ghost model of dark energy in $f(R,T)$ gravity and GR in an anisotropic background. Xu et al. \cite{Xu16} have studied the quantum cosmology and obtained the general form of the quantum potential, gravitational Hamiltonian and the canonical momenta. Shabani and Ziaie \cite{Shabani17} have obtained the late time cosmological solution. In the non-minimal coupled scenario, Sharif and Nawazish \cite{Sharif17} have determined the existence of Noether symmetry and performed the cosmological analysis.

Under $f(R,T)$ gravity, Islam and Basu \cite{Islam18} described the compact star model in the presence of magnetic field and indicated the importance of the solution in the context of interior of compact objects. Elizalde and Khurshudyan \cite{Elizalde18} have discussed the static wormhole with the possible violation of weak and dominant energy conditions. Mishra et al. \cite{Mishra18} have introduced the reconstruction method to obtain an anisotropic cosmological model of the Universe. In metric approach of $f(R,T)$ gravity, Khan et al. \cite{Khan18} have studied the higher dimensional collapse of perfect fluid with spherically symmetric space-time. Wu et al. \cite{Wu18} introduced the non-minimal coupling between the matter and geometry in Palatini formulation by assuming the affine connection and metric as independent field variables. Using metric-affine theories, Barrientos et al. \cite{Barrientos18}  found the field equations and shown its relevance with the field equations of $f(R)$ gravity. Tretyakov \cite{Tretyakov19} introduced the higher derivative matter field in $f(R,T)$ gravity and discussed the stability conditions. Mishra et al. \cite{Mishra19} obtained the bounce solution in an anisotropic space-time. Fisher and Carlson \cite{Fisher19} have re-examined the $f(R,T)$ gravity and claimed that the term $f_2(T)$ should be included in the matter Lagrangian. Tripathy and Mishra \cite{Tripathy20} have presented the phantom cosmological in $f(R,T)$ gravity in different rip cosmologies. Maurya and Tello-Ortiz \cite{Maurya20} have investigated the nature of charged anisotropic fluid spheres and also have shown the similarities and differences between modified gravity and GR. Mishra et al.\cite{Mishra20a} presented the wormhole solution of the model with the square of the trace of the energy momentum tensor. Recently, Tripathy has explored the possibility of Casimir wormholes in an extended gravity theory \cite{Tripathy2020}. Rahaman et al. \cite{Rahaman20} explained the Karmarkar condition in $f(R,T)$ gravity and studied the existence of compact spherical systems whereas with the same condition, Ahmad and Abbas \cite{Ahmad20} have studied the non-adiabatic gravitational collapse in an anisotropic fluid. Mishra and Tripathy \cite{Mishra20b} compared the little rip and specific form of Hubble parameter cosmological models in an anisotropic space-time. Gamonal \cite{Gamonal21} studied with minimal coupling, the slow-roll approximation of cosmic inflation and obtained the modified slow-roll parameters.

The paper is organized as follows: we have presented a brief overview of the $f(R,T)$ gravity and derived the field equations and the basic parameters in Sec II. In Sec III, we have formulated two cosmological models based on quasi de Sitter (QDS) and linear combination of exponential scale factors (LCE). The analysis of different dynamical parameters of the models are performed. Also, we have carried out certain tests of the models through the calculations of the energy conditions and the state finder diagnostic pair. The results and the conclusion are given in Sec IV.

\section{$f(R,T)$ gravity and Basic Parameters}
Harko et al. \cite{Harko11} have proposed the geometrically extended $f(R,T)$ theory of gravity, whose action can be described as,

\begin{equation} \label{eq:1}
S=\int d^4x\sqrt{-g}\left[\frac{1}{16\pi}(f_1(R)+f_2(T))+\mathcal{L}_m\right]. 
\end{equation} 

In above the function $f(R,T)$ has been split into two functions $f_1(R)$ and $f_2(T)$ i.e the function of Ricci scalar $R$ and trace of energy momentum  tensor $T=g^{ij}T_{ij}$ respectively. The Lagrangian has been considered as, $\mathcal{L}_m=-p$. We assumed here, $f_1(R)=R$ and $f_2(T)=2\beta T+2\Lambda_0$, such that $f(R,T)=R+2\beta T+2\Lambda_0$; $\beta$ be the coupling constant and $\Lambda_0$ may be considered as the usual cosmological constant in GR. The matter field considered to be a perfect fluid whose energy momentum tensor can be expressed as, $T_{ij}=(\rho+p)u_iu_j-pg_{ij}$ and $u^iu_i=1$. Here $p$ and $\rho$ respectively denote the pressure and energy density of the cosmic fluid and the trace of the energy momentum tensor is given by $T=\rho-3p$. Now, the field equations of $f(R,T)$ gravity can be derived as \cite{Harko11,Tripathy20},

\begin{equation}\label{eq:2}
R_{ij}-\frac{1}{2}Rg_{ij}=(8\pi+2\beta)T_{ij}+\left[(\rho-p)\beta+\Lambda_0\right]g_{ij}.
\end{equation}
 
When $\beta=0$, eqn. \eqref{eq:2} recovers the field equations of Einstein's GR. It is to note here that $(8\pi+2\beta)$ is the redefined Einstein constant. Due to the geometrical modification incorporated through a minimal matter-geometry coupling, the effective energy momentum tensor $T_{ij}^{EF}$ can be expressed as, 
\begin{equation} \label{eq:3}
T_{ij}^{EF}=\left[\frac{(\rho-p)\beta+\Lambda_0}{8\pi+2\beta}\right]g_{ij}.
\end{equation}
We consider here the Friedmann-Robertson-Walker(FRW) flat space-time,
\begin{equation} \label{eq:4}
ds^{2}=dt^{2}-a^{2}(dx^{2}+dy^{2}+dz^{2}),
\end{equation}
where $a=a(t)$ is the average scale factor. Now, with eqn. \eqref{eq:4}, the field equations \eqref{eq:2} for $f(R,T$) gravity can be obtained as, 

\begin{eqnarray}\label{eq:5}
2\dot{H}+3H^2&=& -\alpha p+\beta \rho+\Lambda_0, \\
3H^2&=&\alpha \rho-\beta p+\Lambda_0. \label{eq:6}
\end{eqnarray}
For brevity, we write $\alpha=8\pi+2\beta$. $H=\frac{\dot{a}}{a}$ is the Hubble parameter. Now, on solving eqns. \eqref{eq:5}-\eqref{eq:6}, we obtain the pressure and energy density of the model as,

\begin{eqnarray}
p&=&-\frac{1}{(\alpha^2-\beta^2)}\left[2\dot{H}\alpha+3H^2(\alpha-\beta)-\Lambda_0(\alpha-\beta)\right], \label{eq:7}\\
\rho&=& \frac{1}{\alpha^2-\beta^2}\left[-2\dot{H}\beta+3H^2(\alpha-\beta)-\Lambda_0(\alpha-\beta)\right]. \label{eq:8}
\end{eqnarray}

The Equation of State (EoS) parameter  $\omega=\frac{p}{\rho}$ can be calculated as,
\begin{eqnarray} \label{eq:9}
\omega=-1+\frac{2(\alpha+\beta)\dot{H}}{2\beta\dot{H}-3H^2(\alpha-\beta)+\Lambda_0(\alpha-\beta)}.
\end{eqnarray}

Another significant study to constrain any cosmological model is the analysis of the energy conditions. Hawking and Ellis \cite{Hawking73} have first formulated the energy conditions.  Basically, the energy conditions are governed by the congruence of space-like, time-like and light-like curves (Hawking and Ellis\cite{Hawking73}, Tahim et al. \cite{Tahim07}), but here we shall constrain to the space-like and time-like curves only. The energy conditions are: (a) $\rho+p\geq 0$, Null energy condition (NEC); (b)  $\rho+p\geq 0$, $\rho\geq 0$, Weak energy condition (WEC); (c) $\rho+3p\geq 0$, Strong energy condition (SEC); and (d) $\rho-p\geq 0$, Dominant energy condition (DEC).  The violation of SEC on cosmological scales has been a point of attention based on the recent observational data in the context of accelerating Universe \cite{Visser97}. In Phantom field models, there is possibility of violation of the NEC. The NEC, WEC, DEC are extensively used in  modified theories of gravity since the energy conditions play a significant role in defining the cosmological evolution, particularly the acceleration or deceleration of cosmic fluid\cite{Atazadeh09,Wang12,Banijmali12,Esmaeili18,Bhatti18,Capozziello18}. The energy conditions for $f(R,T)$ gravity in terms of Hubble parameter can be obtained from eqns. \eqref{eq:7}-\eqref{eq:8} as,

\begin{eqnarray}
\rho+p&=&-\frac{1}{\alpha-\beta}\left[2\dot{H}\right], \nonumber ~~~~ (NEC,WEC) \\
\rho+3p&=& \frac{1}{\alpha^2-\beta^2}\left[-(6\alpha+2\beta)\dot{H}- 6(\alpha-\beta)H^2+2(\alpha-\beta)\Lambda_0\right],\nonumber~~~~ (SEC)\\
\rho-p&=& \frac{1}{\alpha^2-\beta^2}\left[2(\alpha-\beta)\dot{H}+6(\alpha-\beta)H^2-2(\alpha-\beta)\Lambda_0\right].~~~~ (DEC)\label{eq:10}
\end{eqnarray}

It is certain that, a given cosmological model with a specified dynamics decides the energy conditions and all other dynamical parameters such as the pressure, energy density and the EoS parameter. It is also possible to investigate the cosmic dynamics for assumed relationship between the pressure and the energy density. In the present work, we will consider some assumed cosmic dynamics through certain scale factor and then investigate the dynamical parameters of the models.

\section{The Cosmological Models}
In order to understand the background cosmology, we need to incorporate the scale factor to obtain the solution to the field equations. We have studied the cosmological model by incorporating two types of scale factor such as (i))quasi de Sitter (QDS) and (ii) linear combinations of exponential (LCE) functions.

\subsection{Quasi de Sitter model}
The quasi de Sitter scale factor can be expressed in the form $a=e^{{h_{0}t}-\frac{M^{2}t^{2}}{12}}$, where $h_0$ and $M$ are constants. Subsequently the Hubble parameter $H$ can be derived as  $H=\frac{\dot{a}}{a}=h_0-\frac{M^2t}{6}$. For this model, the slope of the Hubble parameter becomes $\dot{H}=-\frac{M^2}{6}$. One should note that, this model reduces to the de Sitter model for $M=0$.

The deceleration parameter, scalar expansion and the spatial volume for this QDS model are obtained as,  
\begin{eqnarray}
q=-1+\frac{6M^2}{(6h_0-M^2t)^2}; ~~~~~~~\theta =3H=3\left(h_0-\frac{M^2t}{6}\right);~~~~~~~ V=a^{3}=e^{3\left(h_0t-\frac{M^2t^2}{12}\right)}.
\end{eqnarray}

\begin{figure}[!htp]
\centering
\includegraphics[scale=0.50]{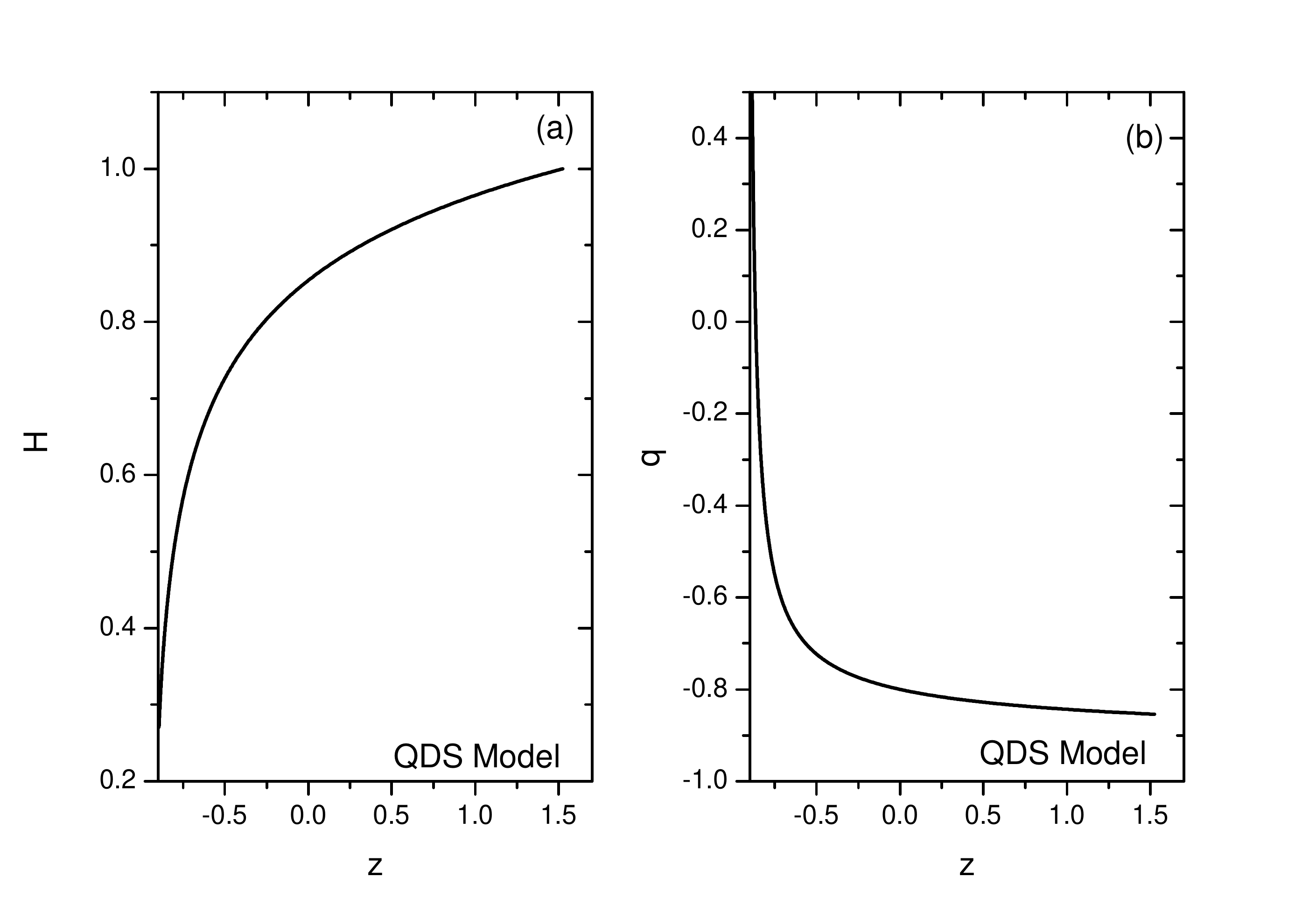}
\caption{(a)Graphical behavior of the Hubble parameter $H$ as a function of redshift  and (b) behaviour of deceleration parametr $q$.} 
\label{Fig1}
\end{figure}

We show the behaviour of the QDS model in FIG.1(a) and that of the deceleration parameter in FIG.1(b). The Hubble parameter decreases with cosmic expansion where as the deceleration parameter is observed to increase with cosmic evolution. The parameters of the QDS model obviously affect the behaviour of the Hubble parameter and the deceleration parameter. We have considered suitable parameter space for the QDS model so as to obtain the observational deceleration parameter at the present epoch. We chose $h_0=1$ and $M=-\sqrt{\frac{3}{2}}(\sqrt{5}+3)$ so that, the deceleration parameter at the present epoch becomes $q_0=q(z=0)=-0.8$. It is worth to mention here that a recent analysis constrained the deceleration parameter at the present epoch as $q_0=-1.08\pm 0.29$ \cite{Camarena2020}. In another work, Goswami et al. have obtained a constraint as $q_0=-0.59$ \cite{Goswami2021}. In this sense, the QDS model is consistent with the present cosmic scenario.

Incorporating the QDS scale factor in \eqref{eq:7}-\eqref{eq:8}, we get the pressure and energy density as

\begin{eqnarray}
p&=&-\frac{1}{(\alpha^2-\beta^2)}\left[-\frac{M^2}{3}\alpha+3\left(h_0-\frac{M^2t}{6}\right)^2(\alpha-\beta)-\Lambda_0(\alpha-\beta)\right], \label{eq:11}\\
\rho&=& \frac{1}{(\alpha^2-\beta^2)}\left[\frac{M^2}{3}\beta+3\left(h_0-\frac{M^2t}{6}\right)^2(\alpha-\beta)-\Lambda_0(\alpha-\beta)\right].\label{eq:12}
\end{eqnarray}

Subsequently the equation of state (EoS) parameter $\omega=\frac{p}{\rho}$ can be calculated as,
\begin{equation}\label{eq:13}
\omega=-1+\frac{M^2(\alpha+\beta)}{M^2\beta+9\left(h_0-\frac{M^2t}{6}\right)^2(\alpha-\beta)-3\Lambda_0(\alpha-\beta)}.
\end{equation}

\begin{figure}[!htp]
\centering
\includegraphics[scale=0.3]{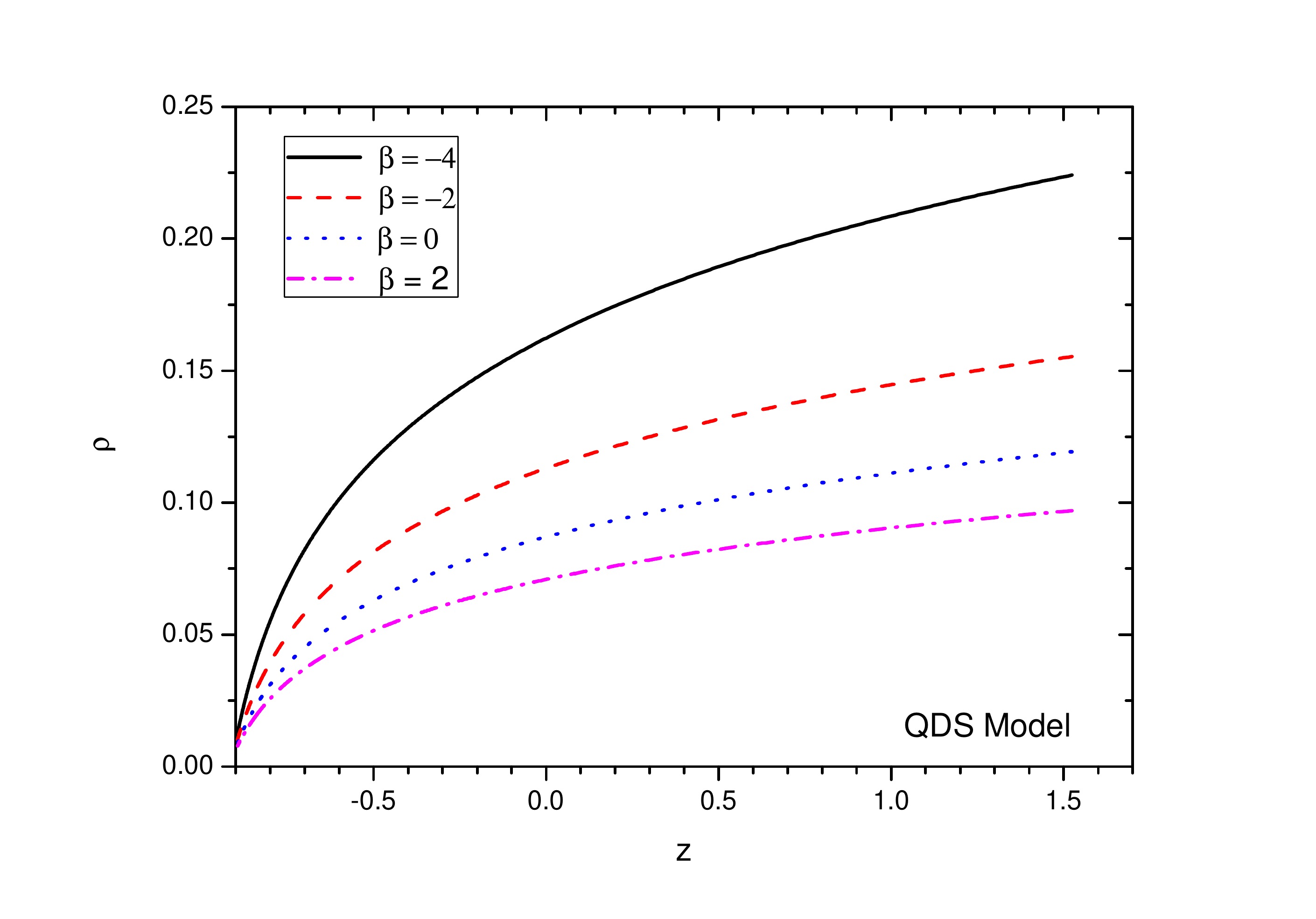}
\includegraphics[scale=0.3]{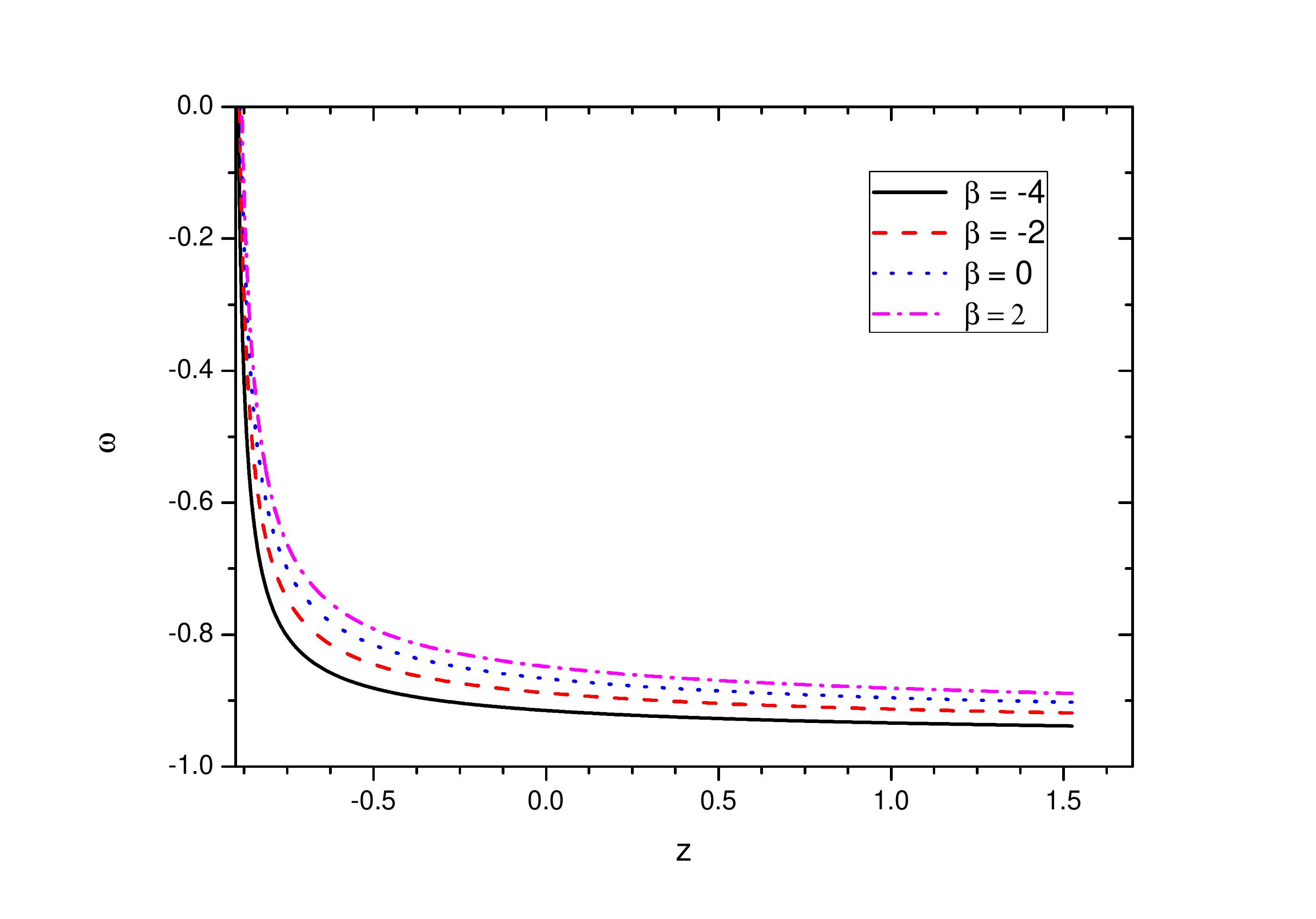}
\caption{Graphical behavior of the energy density $\rho$ (left panel) and the EoS parameter $\omega$ (right panel) as functions of redshift for four representative values of the coupling constant $\beta$ namely $\beta=-4, -2, 0$ and 2. Here we have assumed the parameter space $\Lambda=0.001, h_0=1$.}
\label{Fig1b}
\end{figure}
In the left panel of FIG. 2, the behaviour of the energy density with respect to cosmic redshift is shown. We have defined the redshift as $z=-1+\frac{a_0}{a}$ where $a_0$ is the scale factor at the present epoch. We have chosen four representative values of the coupling constant $\beta$ namely $\beta=-4,-2,0$ and 2. The parameter $\alpha$ can be calculated from the assumed values of the coupling constant. The coupling constant values are chosen so that, the energy density remains positive through the cosmic evolution. It should be recalled here that, the QDS model with $\beta=0$ recovers the GR features. For a given value of the coupling constant, the energy density decreases from some higher value to negligible values at late times. At a given epoch, the value of the energy density depends on the choice of the coupling parameter in such a manner that, with an increase in the value of $\beta$, the energy density decreases. In the right panel of FIG.2, the dynamical behaviour of the EoS parameter is shown as a function of the redshift. It is observed that, the equation of state parameter for a given coupling constant parameter increases from a higher negative values at an early epoch to low negative values at late times. The role of the coupling constant $\beta$ is quite visible at an early epoch than at late epoch. With an increase in $\beta$ value, the EoS parameter is found to increase. From our model, we have obtained the EoS parameter at the present epoch as $-0.915, -0.888, -0.867$ and $-0.848$ corresponding to the $\beta$ values $-4, -2, 0$ and 2 respectively. The observational evidences suggest the range of $\omega$ as, $\omega=-1.035^{+0.055}_{-0.059}$ \cite{Amanullah10}; $\omega=-1.073^{+0.090}_{-0.089}$\cite{Hinshaw13}; $\omega=-1.06^{+0.11}_{-0.13}$ \cite{Kumar14} and the most recent Planck collaboration suggests $\omega=-1.03\pm 0.03$ \cite{Aghanim20}. The result obtained for the EoS parameter are in accordance with the observational results.

The corresponding energy conditions for the QDS scenario can be derived as, 
\begin{eqnarray}
\rho+p &=&\frac{1}{(\alpha-\beta)}\left[\frac{M^2}{3}\right], \nonumber \\ 
\rho+3p &=&\frac{1}{(\alpha^2-\beta^2)}\left[(3\alpha+\beta)\frac{M^2}{3}-6(\alpha-\beta)\left(h_0-\frac{M^2t}{6}\right)^2+2(\alpha-\beta)\Lambda_0\right], \nonumber \\
\rho-p&=&-\frac{1}{(\alpha^2-\beta^2)}\left[(\alpha-\beta)\frac{M^2}{3}-6(\alpha-\beta)\left(h_0-\frac{M^2t}{6}\right)^2+2(\alpha-\beta)\Lambda_0\right].\label{eq:14}
\end{eqnarray}
The NEC for the present $f(R,T)$ gravity model depends on the slope of the Hubble parameter and we have a constant slope in case of the QDS model. Therefore, the NEC in the QDS scenario remains positive but non-evolving. In other words, the present QDS scenario favours the satisfaction of the NEC. On the other hand, the $f(R,T)$ gravity suggests the violation of SEC in view of the accelerated expansion of the Universe. In FIG. 3, we show the behaviour of the energy conditions for the QDS model which clearly shows the violation of SEC and non-violation of DEC and NEC. For the de Sitter scenario with $M=0$, the NEC becomes zero and the other energy conditions SEC and DEC remain independent of time and  throughout the evolution, the behaviour of the energy conditions would not change.

\begin{figure}[!htp]
\centering
\includegraphics[scale=0.50]{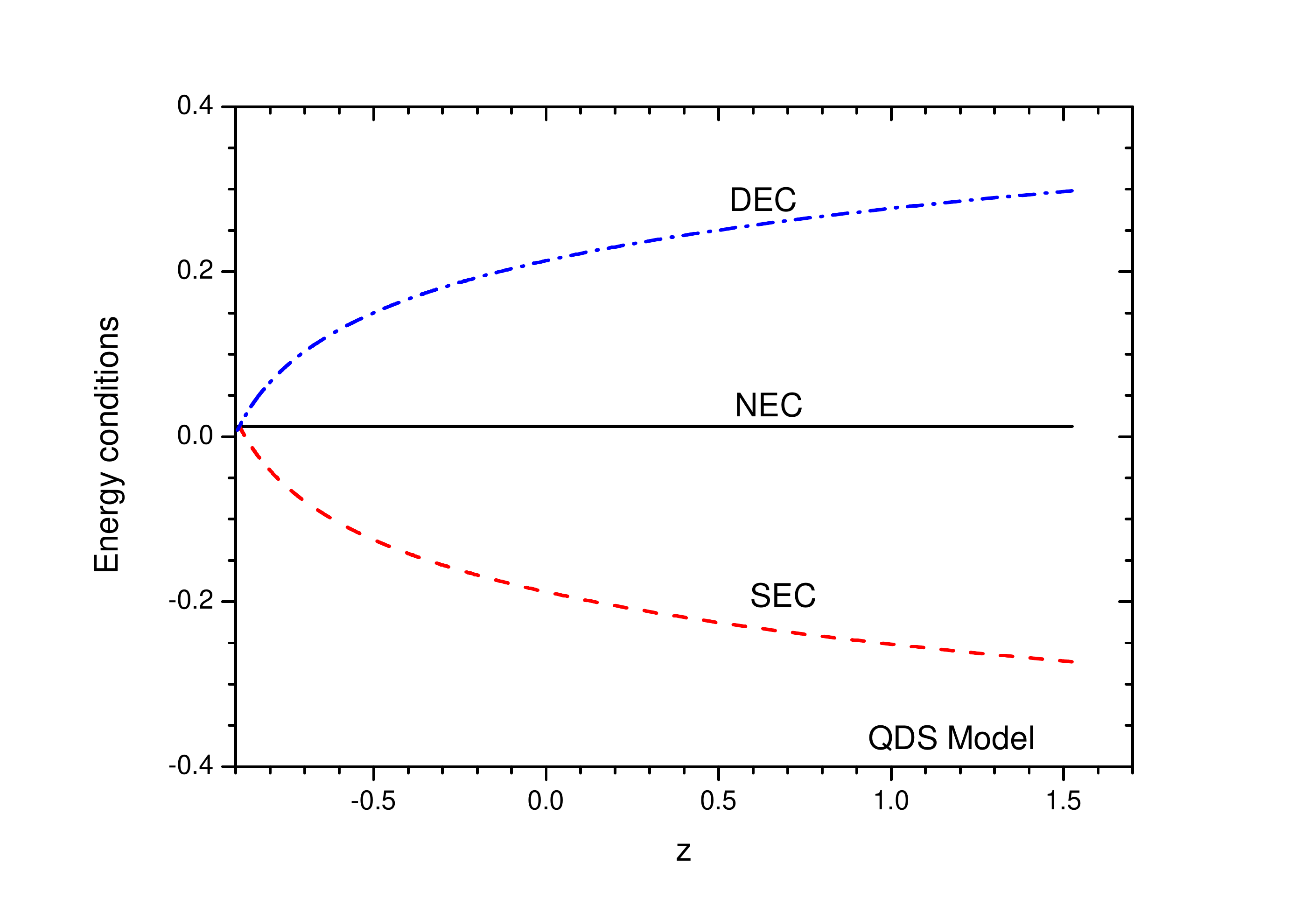}
\caption{Graphical behaviour of the energy conditions versus  redshift for the parametric values $\beta=-2, h_0=1$.}
\label{Fig2}
\end{figure}

It should be interesting to discuss the role played by the modified gravity in the context of an accelerating cosmic scenario. Usually in GR, the late time cosmic speed up phenomena is associated with an exotic dark energy form. However, in the purview of modified extended gravity model, the accelerating aspect is handled by an effective energy-momentum tensor as given by eqn. \eqref{eq:3}. Let us now denote this effective energy momentum tensor as $T_{ij}^{EF}=\Lambda_{eff}g_{ij}$, so that the effective cosmological constant in the extended gravity model becomes
\begin{equation}
\Lambda_{eff}=\frac{\left(\rho-p\right)\beta+\Lambda_0}{\alpha}.
\end{equation}
Consequently, the density parameter corresponding to this pseudo vacuum label becomes
\begin{equation}
\Omega_{\Lambda}=\frac{\alpha\Lambda_{eff}}{3H^2}.
\end{equation}

In FIG.4(a) and (b), we display the evolutionary aspects of the effective cosmological constant and the density parameter respectively for four representative values of the coupling constant $\beta$. The effective cosmological constant becomes positive for positive values of $\beta$ and assumes negative values for negative $\beta$. As we have already mentioned earlier, for $\beta=0$, the effective cosmological constant reduces to the usual cosmological constant in GR. While the effective cosmological constant increases with the cosmic expansion for negative values of  $\beta$, for positive value of the coupling constant it decreases with cosmic expansion. For a given $\beta$, the density parameter $\Omega_{\Lambda}$ as obtained from the present QDS model remains almost flat for a substantial time zone during the cosmic evolution. However, the density parameter is found to decrease from a higher positive value to lower positive values for positive $\beta$. On the other hand, for negative values of the $\beta$, $\Omega_{\Lambda}$ increases from higher negative values to lower negative values.
\begin{figure}[!htp]
\centering
\includegraphics[scale=0.50]{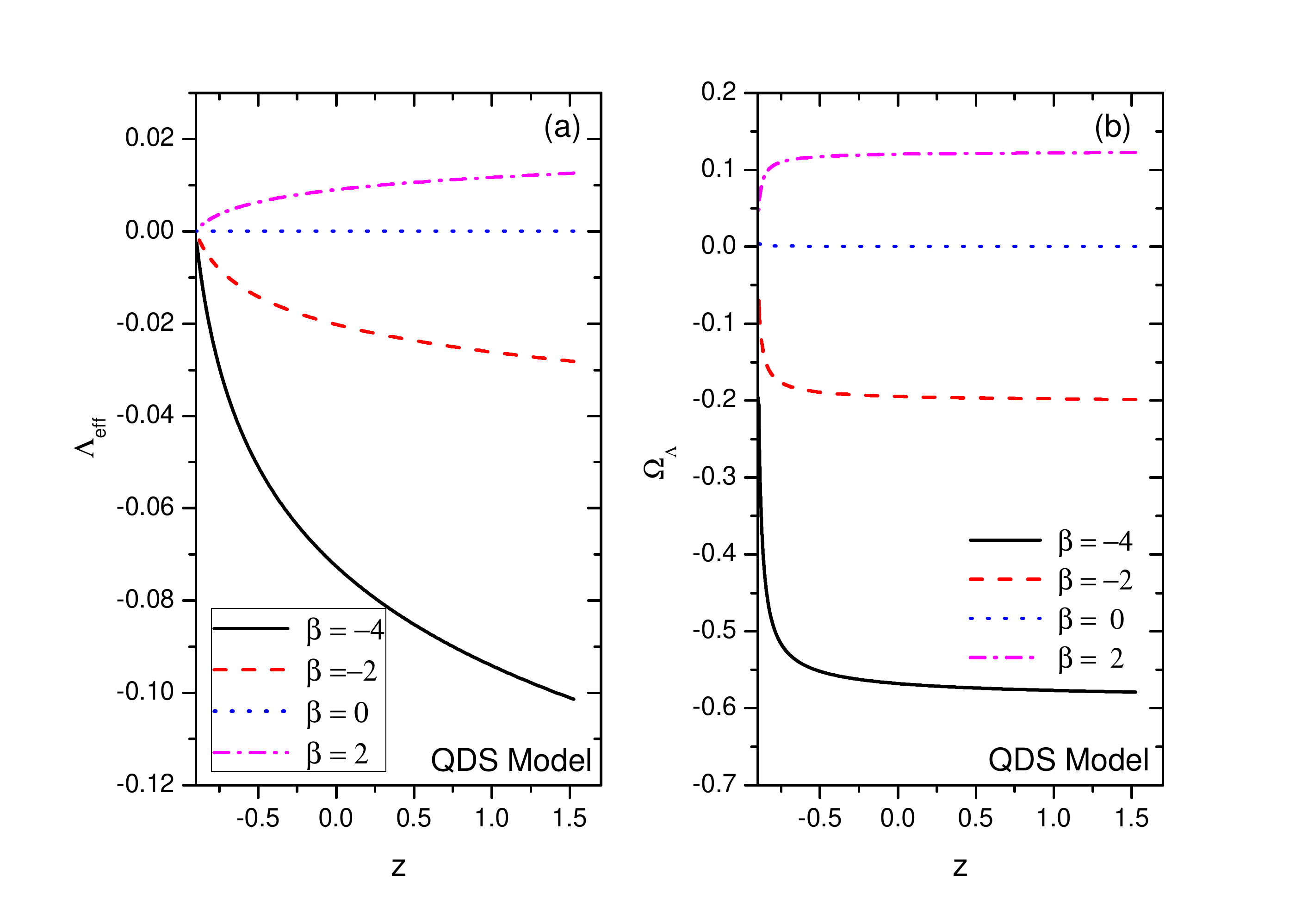}
\caption{(a)Graphical behavior of $\Lambda_{eff}$ and (b) the density parameter $\Omega_{\Lambda}$  as functions of redshift.}
\label{Fig3a}
\end{figure}

The geometrical diagnostics state finder pair $(j,s)$, where $j$ be the jerk parameter and $s$ be the snap parameter can be obtained for the QDS model as,
\begin{eqnarray}
j=\frac{\dddot{a}}{aH^3}=1-\frac{18M^2}{(6h_0-M^2t)^2};~~~~~~~s=\frac{a^{(4)}}{aH^4}=\frac{4M^2}{(6h_0-M^2t)^2-4M^2}. \label{eqn.24}
\end{eqnarray}

We can observe here that for $M=0$, the diagnostic pair $(j,s)$ reduces to $(1,0)$ irrespective of the value of the parameter $h_0$; so also the deceleration parameter becomes $-1$ and the scale factor reduces to a de-Sitter function. Therefore, although we understand that the model would not behave as $\Lambda$CDM model, but we are compelled to assume the representative value $M\neq 0$ for the model. Another argument is that in the expression of $j$ in Eq. \eqref{eqn.24}, the second term provides a positive value and therefore, the model does not coincide with  the concordant value  $j_0=1$. In FIG.5(a) we show the dynamical behaviour of the jerk parameter as a function of redshift which provides a decreasing trend of $j$ with the expansion of the Universe.  At the present epoch ($z=0$), the QDS model within the chosen parameter space, predicts $j_0=0.4$ and shows a possible departure from the concordance $\Lambda$CDM value $j_0=1$. The present result is quite consistent with the results obtained in a recent analysis \cite{Goswami2021}. In FIG.5(b), the dynamical aspects of the snap parameter, $s$ is shown. It may be observed that, for a substantial time zone, the snap parameter remains constant and then it displays a quick jump followed by a sharp drop at a late epoch. At the present epoch, the present model predicts $s_0=-0.00085$ which also signals a small departure from the concordant $\Lambda$CDM value $s_0=0$.  \\
\begin{figure}[!htp]
\centering
\includegraphics[scale=0.50]{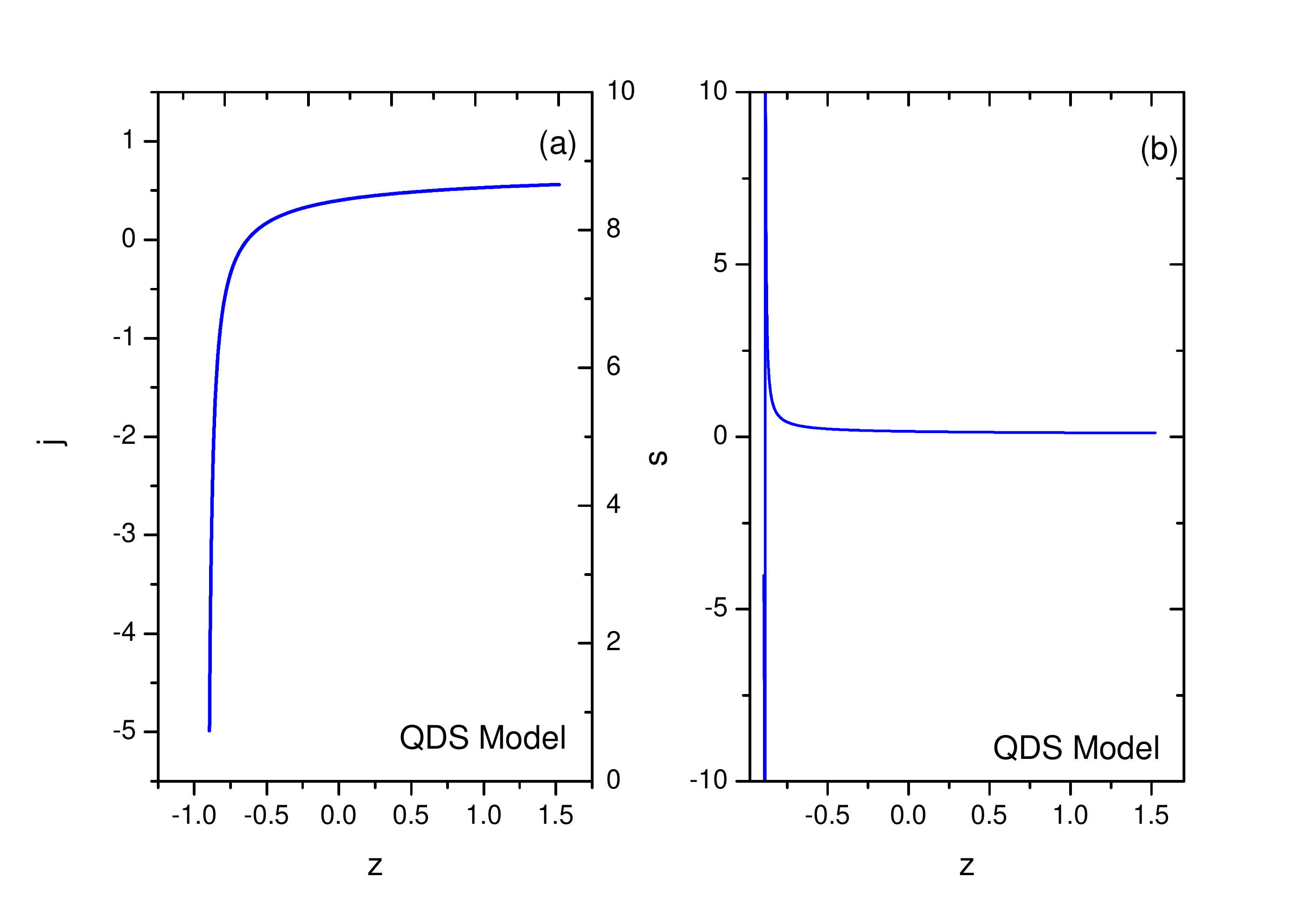}
\caption{(a)Graphical behavior of $j$ and (b) $s$  as functions of redshift.}
\label{Fig3}
\end{figure}

%
%

Now, the throat radius of Morris-Thorne wormhole can be calculated for the QDS model using the relation \cite{Babichev04,Astashenok12,Tripathy20}

\begin{eqnarray} \label{eqn.26}
\dot{R}=CR^2(\rho+p),
\end{eqnarray}  
where $C$ be the dimensionless positive constant. On integration, we get,
\begin{equation}
\frac{1}{R}=\frac{C}{(\alpha-\beta)}\frac{M^2}{3}t+C_1,
\end{equation}
where $C_1$ be the integration constant. It is possible to experience the big trip ($t=t_B$) when $C_1=-\frac{C}{(\alpha-\beta)}\frac{M^2}{3}t_B$. So the wormhole throat radius can be obtained as, $R=\frac{3(\alpha-\beta)}{CM^2(t_B-t)}$. If $R_0$ be the wormhole throat radius at $t=t_0$, then we can have the big trip at $t_B=t_0+\frac{3(\alpha-\beta)}{R_0CM^2}$.
\subsection{Linear combination of exponential factor model}
Another important scale factor to study the background cosmology is the LCE scale factor which can be described as 
$a(t)=\sigma e^{\lambda t}+\tau e^{-\lambda t} $.
Unlike in exponential scale factor, this scale factor consists of both positive and negative exponential factors. The significance of the scale factor is that when $\sigma=0$, the scale factor reduces to negative exponential factor whereas for $\tau=0$, it reduces to a de Sitter scenario. Such a scale has been used earlier to study a bouncing scenario \cite{Tripathy2021, Bamba2014}. The Hubble parameter $H$ for this LCE model can be obtained as 
\begin{equation}
H=\dot{\frac{a}{a}}=\frac{\sigma\lambda e^{\lambda t}-\tau\lambda e^{-\lambda t}}{\sigma e^{\lambda t}+\tau e^{-\lambda t}}
\end{equation}
and subsequently its slope becomes
\begin{equation}
\dot{H}=\frac{\ddot{a}}{a}-\frac{\dot{a}^{2}}{a^{2}}=\frac{4\sigma\tau\lambda^{2}}{(\sigma e^{\lambda t}+\tau e^{-\lambda t})^{2}}.
\end{equation}

The deceleration parameter, scalar expansion and the volume of the LCE model can be obtained as, 
\begin{eqnarray}
q=-1-\frac{4\sigma \tau \lambda^2}{(\sigma e^{\lambda t}-\tau e^{-\lambda t})^2
};~~~~~~~ \theta =3\left(\frac{\sigma\lambda e^{\lambda t}-\tau\lambda e^{-\lambda t}}{\sigma e^{\lambda t}+\tau e^{-\lambda t}}\right); ~~~~~~~V=\left(\sigma e^{\lambda t}+\tau e^{-\lambda t}\right)^{3}.
\end{eqnarray}
\begin{figure}[!htp]
\centering
\includegraphics[scale=0.50]{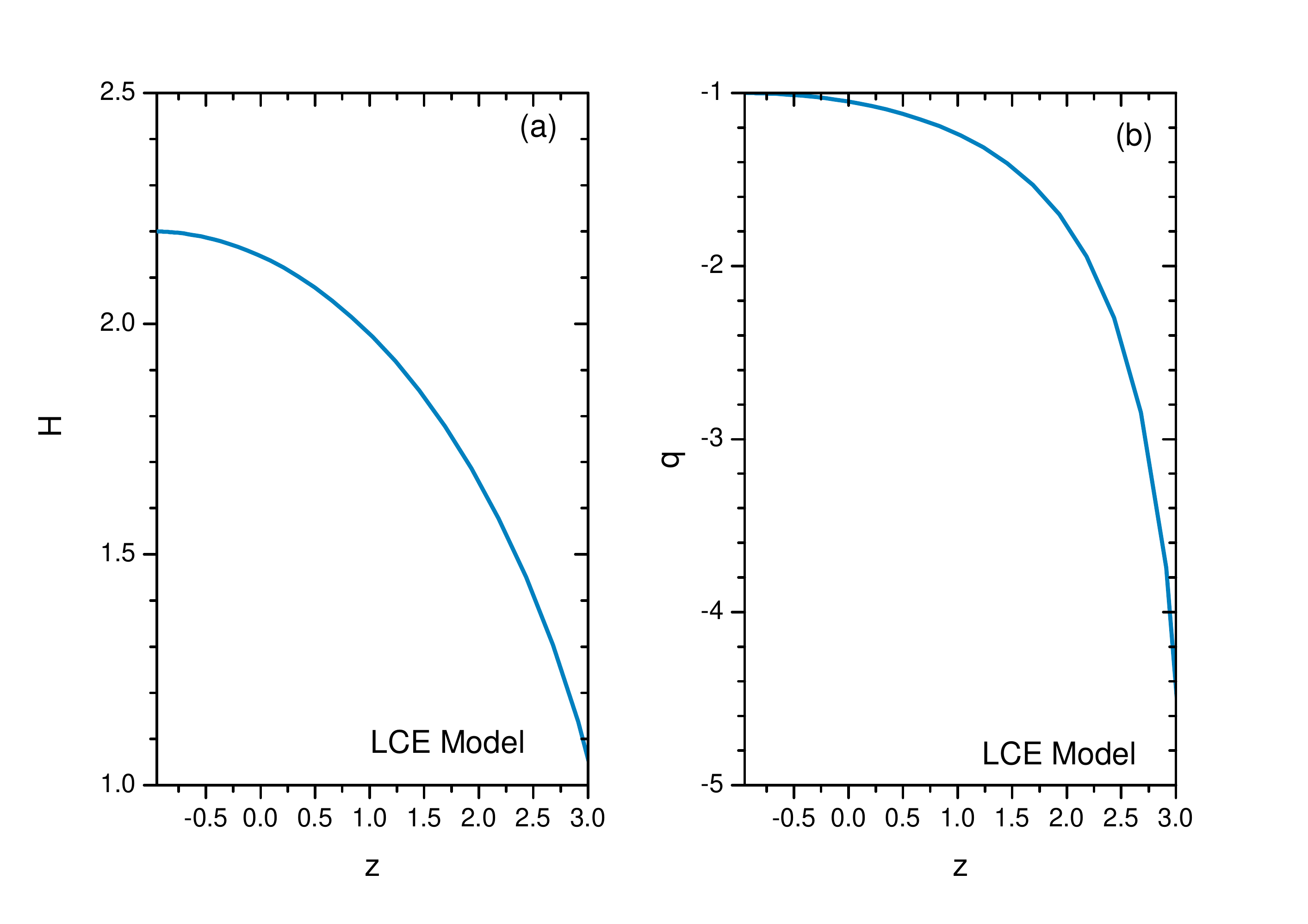}
\caption{(a)Graphical behavior of the Hubble parameter $H$ as a function of redshift  and (b) behaviour of deceleration parameter $q$.} 
\label{Fig1a}
\end{figure}
In FIG.6(a) and (b), we show the evolutionary aspect of the Hubble parameter and the deceleration parameter respectively. Both the Hubble parameter and the deceleration show an increasing trend  with the cosmic expansion. In order to plot the figures, we have followed the earlier work in Ref. \cite{Tripathy2021} and considered $\sigma=\tau=1$ and $\lambda=2.2$. The deceleration parameter increases from  higher negative values to become $q=-1$ at late epoch. At the present epoch, the present LCE model predicts the deceleration parameter to be $q_0=-1.048$ which is quite close to the earlier estimates and therefore compatible with observations.

For the $f(R,T)$ field equations in FRW space-time, the pressure and the energy density can be obtained as

\begin{eqnarray}
p&=&-\frac{1}{(\alpha^2-\beta^2)}\left[2\left(\frac{4\sigma\tau\lambda^{2}}{(\sigma e^{\lambda t}+\tau e^{-\lambda t})^{2}}\right)\alpha+3\left(\frac{\sigma\lambda e^{\lambda t}-\tau\lambda e^{-\lambda t}}{\sigma e^{\lambda t}+\tau e^{-\lambda t}}\right)^2(\alpha-\beta)-\Lambda_0(\alpha-\beta)\right],\\
\rho&=& \frac{1}{\alpha^2-\beta^2}\left[-2\left(\frac{4\sigma\tau\lambda^{2}}{(\sigma e^{\lambda t}+\tau e^{-\lambda t})^{2}}\right)\beta+3\left(\frac{\sigma\lambda e^{\lambda t}-\tau\lambda e^{-\lambda t}}{\sigma e^{\lambda t}+\tau e^{-\lambda t}}\right)^2(\alpha-\beta)-\Lambda_0(\alpha-\beta)\right].
\end{eqnarray}
Using the above expressions for pressure and the energy density, we obtain the EoS parameter as
\begin{eqnarray}
\omega=-1+\frac{2(\alpha+\beta)(4\sigma \tau \lambda^2)}{2\beta(4\sigma \tau \lambda^2)-3(\alpha-\beta)(\sigma \lambda e^{\lambda t}-\tau \lambda e^{-\lambda t})^2+\Lambda_0(\alpha-\beta)(\sigma e^{\lambda t}+\tau e^{-\lambda t})^2}.
\end{eqnarray}

\begin{figure}[!htp]
\centering
\includegraphics[scale=0.50]{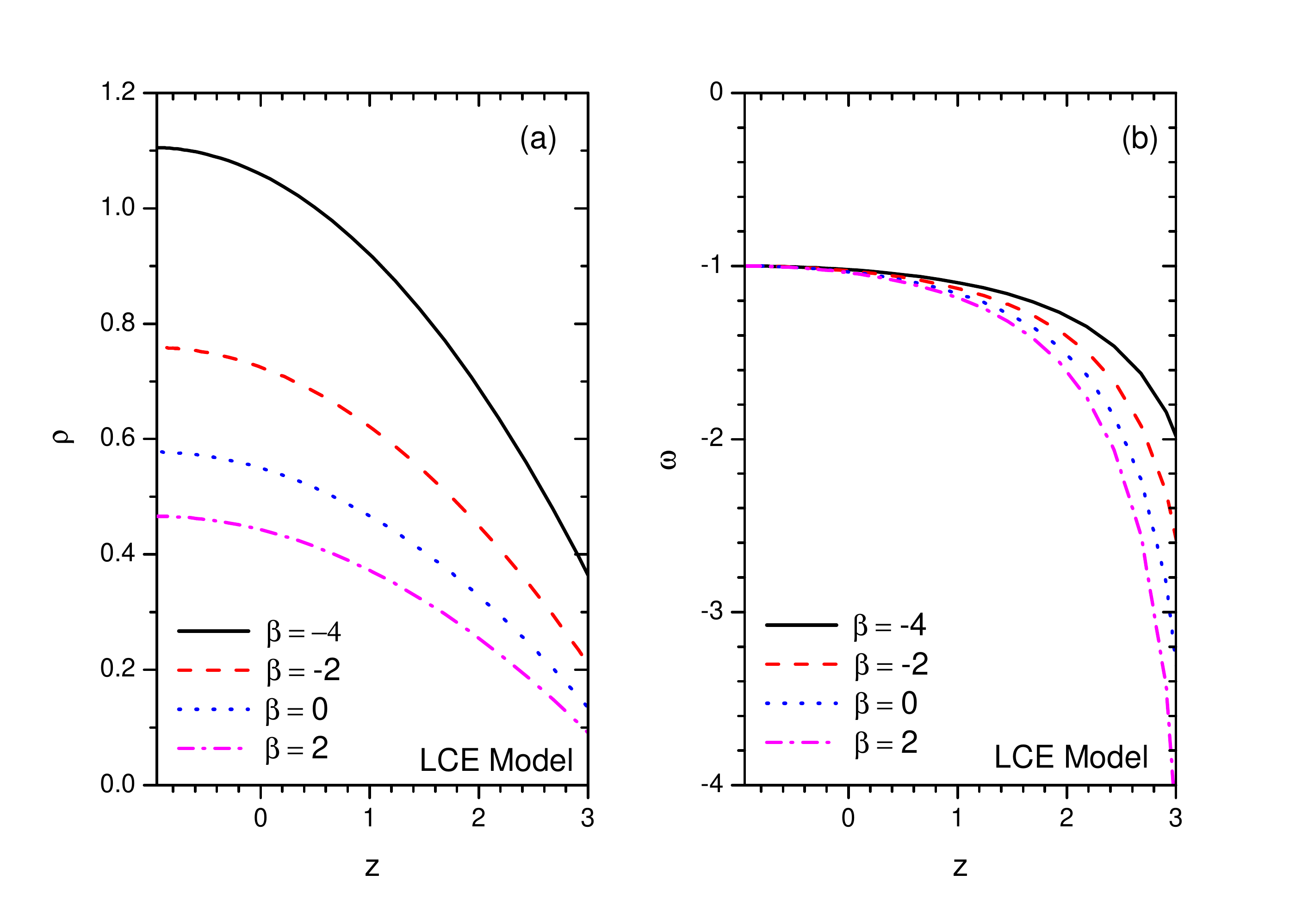}
\caption{(a) Graphical behavior of the energy density $\rho$ and (b) the EoS parameter $\omega$ as functions of redshift for four representative values of the coupling constant $\beta$ namely $\beta=-4, -2, 0$ and 2. Here we have assumed the parameter space $\lambda=2.2$.}
\label{Fig1}
\end{figure}
The left and right panel of FIG. 7 respectively show the behaviour of the energy density and the EoS parameter for the LCE model. Here we have considered four representative values of the coupling constant $\beta$ so as to get positive energy density. The exponential factor $\lambda$ appearing in the scale factor is assumed to be 2.2. The energy density changes its behaviour based on the model parameter $\beta$. In general, for a given value of $\beta$, the energy density is found to increase with the cosmic evolution. At a given epoch, with an increase in $\beta$, the energy density decreases.  Similar behaviour is displayed by the EoS parameter. While $\omega$ increases with the cosmic evolution for a given $\beta$, it decreases with an increase in $\beta$ at a given epoch. However, at late times of cosmic evolution, the EoS parameter of the LCE model converges to the concordant $\Lambda$CDM value $\omega=-1$ irrespective of the values the coupling constant.  The behaviour the EoS parameter shows that, the present LCE model is dominated by a phantom field phase. The LCE model predicts the EoS parameter at the present epoch as $\omega=-1.02, -1.026, -1.032, -1.037$ corresponding to the respective $\beta$ values $-4,-2,0$ and 2. These predicted EoS parameter values are quite compatible with that suggested by Planck results \cite{Aghanim20} and that of other recent works \cite{Amanullah10,Hinshaw13,Kumar14}.

Usually energy conditions put some restrictions on theoretical models. The energy conditions \eqref{eq:10} for the LCE model may be expressed as,
\begin{eqnarray}
\rho+p &=&- \frac{2}{(\alpha-\beta)}\left[\frac{4\sigma \tau \lambda^{2}}{(\sigma e^{\lambda t}+\tau e^{-\lambda t})^2}\right], \\
\rho+3p &=& \frac{1}{(\alpha^2-\beta^2)}\left[-(6\alpha+2\beta)\frac{4\sigma\tau\lambda^{2}}{(\sigma e^{\lambda t}+\tau e^{-\lambda t})^{2}}-6(\alpha-\beta)\frac{(\sigma\lambda e^{\lambda t}-\tau\lambda e^{-\lambda t})^2}{(\sigma e^{\lambda t}+\tau e^{-\lambda t})^2}+2(\alpha-\beta)\Lambda_0\right],\\
\rho-p &=&\frac{1}{(\alpha^2-\beta^2)}\left[2(\alpha-\beta)\frac{4\sigma\tau\lambda^{2}}{(\sigma e^{\lambda t}+\tau e^{-\lambda t})^{2}}+6(\alpha-\beta)\frac{(\sigma\lambda e^{\lambda t}-\tau\lambda e^{-\lambda t})^2}{(\sigma e^{\lambda t}+\tau e^{-\lambda t})^2} -2(\alpha-\beta)\Lambda_0\right].
\end{eqnarray}

\begin{figure}[!htp]
\centering
\includegraphics[scale=0.5]{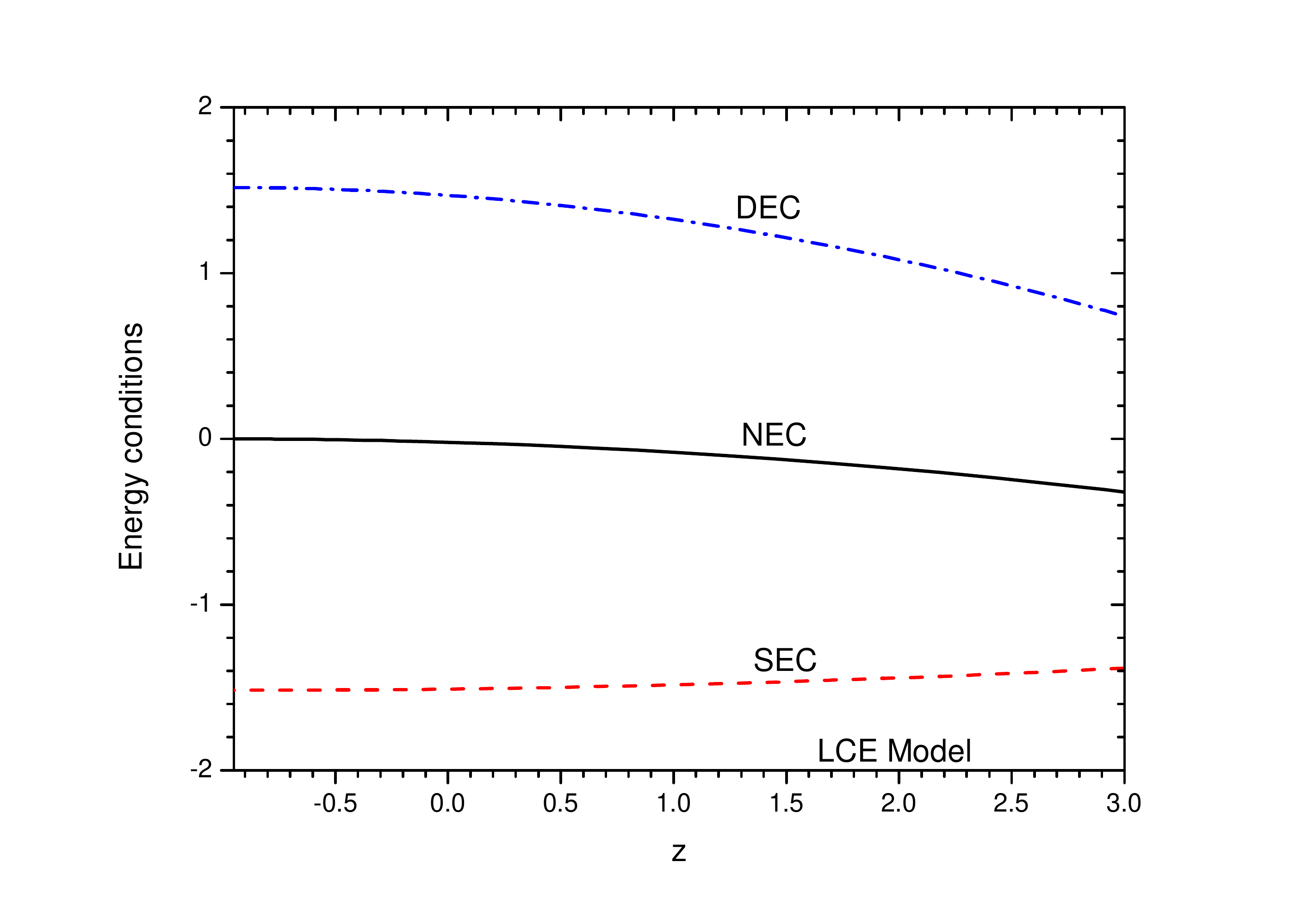}
\caption{Graphical behaviour of the energy conditions versus  redshift for the parametric values $\beta=-2, \lambda=2.2$.}
\label{Fig2b}
\end{figure}
The energy conditions for the LCE model are displayed in FIG.8. For the plots, we have considered only one value of the coupling constant i.e $\beta=-2$. It is obvious from the figure that, the LCE model within the purview of the extended gravity theory violates the NEC and SEC. However, the DEC is satisfied throughout the cosmic evolution. During the discussion of the dynamical behaviour of the EoS parameter, we have inferred that, the present model favours a phantom field like behaviour. It should be emphasized here that, for phantom models, the NEC is usually violated. In fact, the same has happened in the present case. 

\begin{figure}[!htp]
\centering
\includegraphics[scale=0.50]{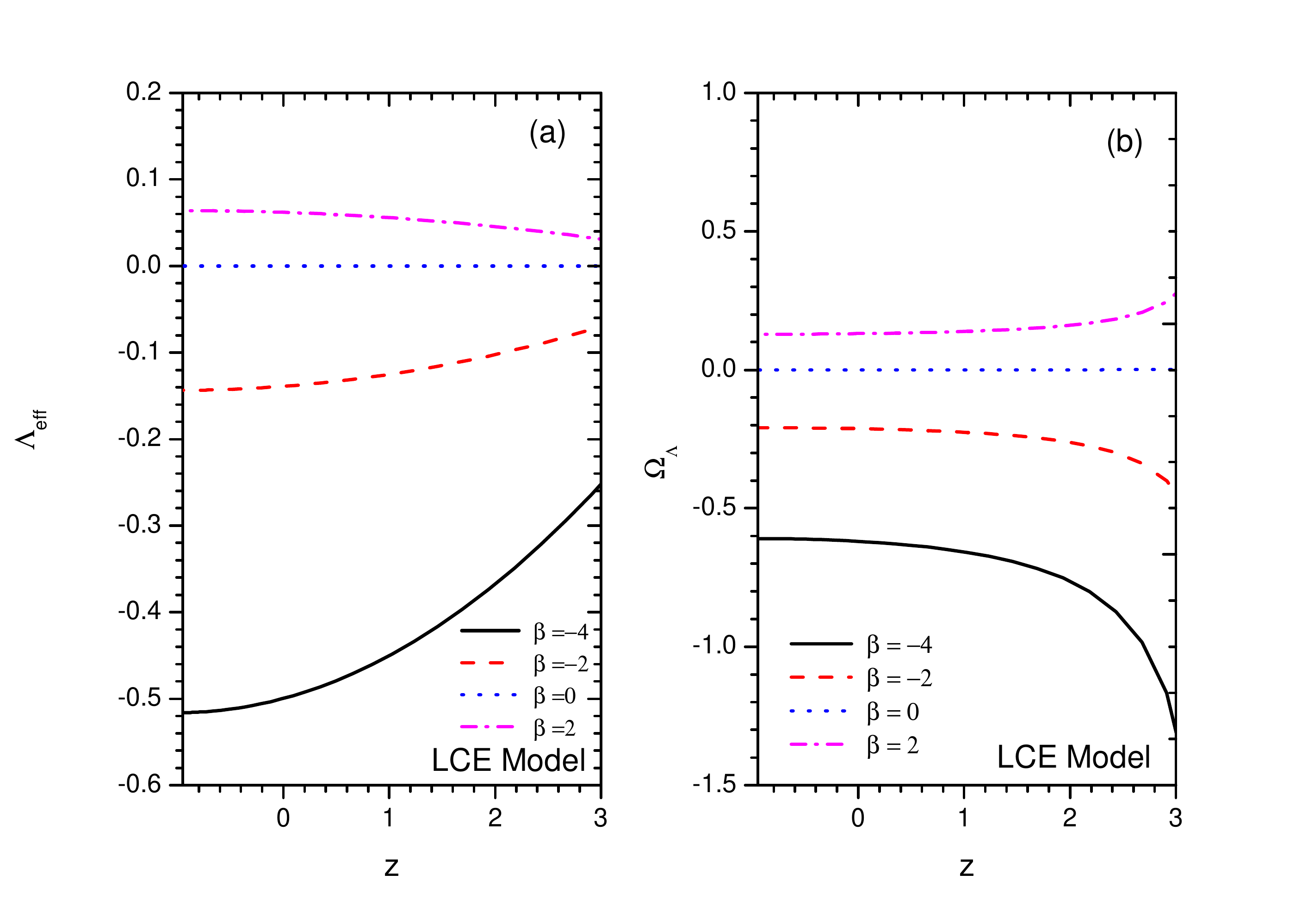}
\caption{(a)Graphical behavior of $\Lambda_{eff}$ and (b) the density parameter $\Omega_{\Lambda}$  as functions of redshift.}
\label{Fig3b}
\end{figure}
As before, for the present model we wish to present the dynamical aspects of the effective cosmological constant $\Lambda_{eff}$ and the corresponding density parameter $\Omega_{\Lambda}$ in FIG.9(a) and (b) respectively. The effective cosmological constant becomes a positive quantity for positive values of $\beta$ and negative for negative $\beta$ and reduces to the usual cosmological constant $\Lambda_0$ for $\beta=0$. For negative values of $\beta$, $\Lambda_{eff}$ decreases from a low negative value at an early epoch to higher negative values at late times. The decrement in $\Lambda_{eff}$ is faster for low values of $\beta$. On the other hand, for positive $\beta$, $\Lambda_{eff}$ increases slowly within the positive domain. The density parameter $\Omega_{\Lambda}$ also assumes positive values for positive $\beta$ and negative values for negative $\beta$. While $\Omega_{\Lambda}$ increases along with the cosmic expansion for $\beta<0$, it increases for $\beta>0$.

The state finder diagnostic pair $(j,s)$ for the LCE scenario are calculated as, 

\begin{eqnarray}
j=&\frac{\ddot{H}}{H^{3}}-(2+3q)=\left(\frac{\sigma e^{\lambda t}+\tau e^{-\lambda t}}{\sigma e^{\lambda t}-\tau e^{-\lambda t}}\right)^{2}; ~~~~~~~s=\frac{r-1}{3(q-0.5)}=\frac{-8\sigma\tau}{3(\sigma e^{\lambda t}+\tau e^{-\lambda t})^{2}}.
\end{eqnarray}
In FIG. 10, we show the evolutionary aspect of the state finder pair. While the jerk parameter is observed to decrease with cosmic evolution, the snap parameter increases. At late time of cosmic evolution, both the parameter converge to the concordant $\Lambda$CDM values $\{1,0\}$. In this sense, we may infer that, the LCE model reduces to the $\Lambda$CDM model at late  times. At the present epoch, the predicted values of the statefinder pair are $\{1.048, -0.0104\}$ which are close to the concordant $\Lambda$CDM values $\{1,0\}$.
\begin{figure}[!htp]
\centering
\includegraphics[scale=0.50]{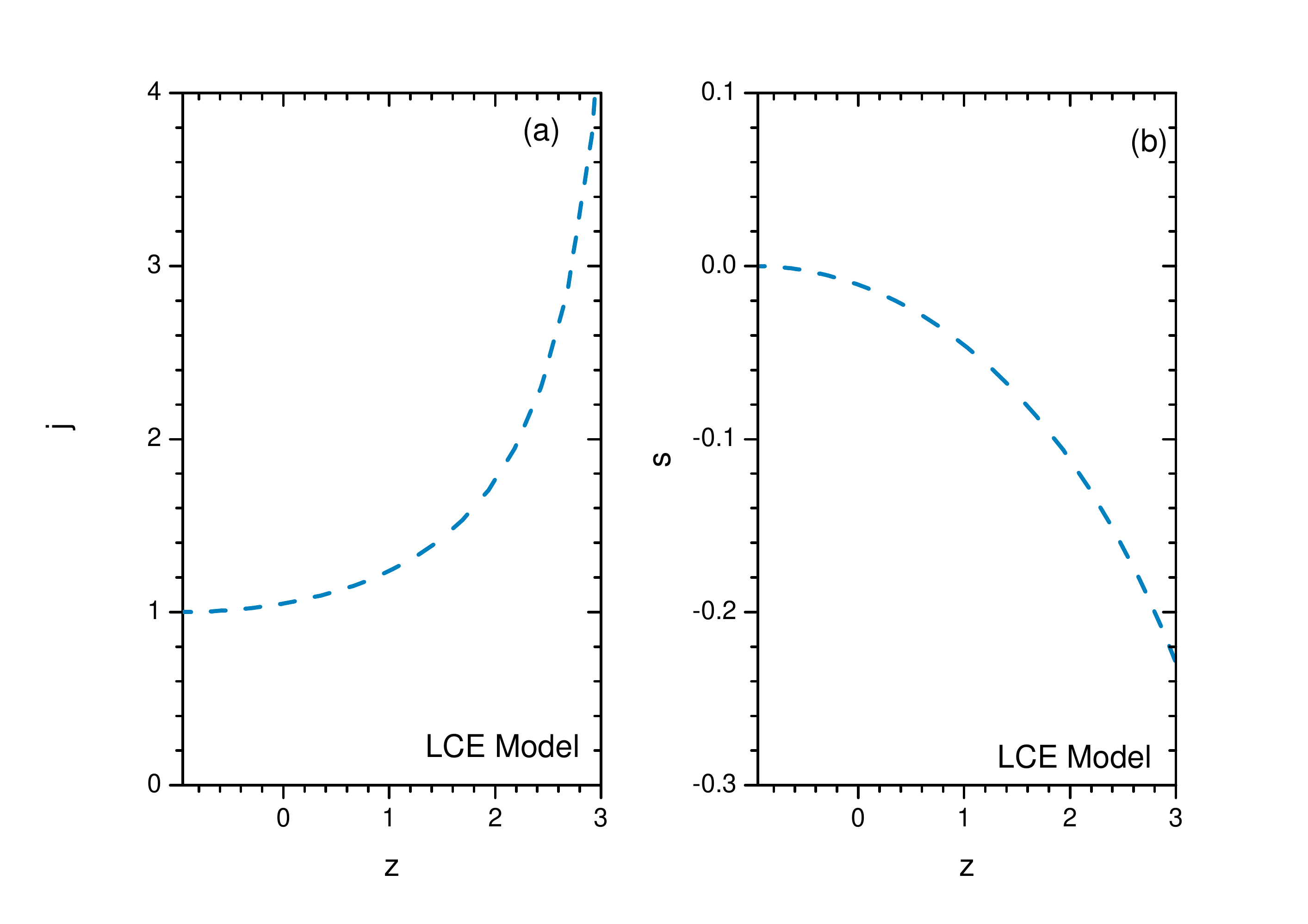}
\caption{(a)Graphical behavior of $j$ and (b) $s$  as functions of redshift.}
\label{Fig3}
\end{figure}


%
%

Following the procedure mentioned in the QDS case, the throat radius for this model is obtained as,
\begin{equation}\label{eqn.39}
R(t)=\frac{(\alpha+\beta)(\sigma e^{\lambda t}+\tau e^{-\lambda t})}{2C\lambda(\sigma e^{\lambda t}-\tau e^{-\lambda t})}.
\end{equation}

It can be noted that when $t\rightarrow 0$, then $R\rightarrow \frac{(\alpha+\beta)(\sigma+\tau)}{2C\lambda(\sigma-\tau)}$ whereas when $t\rightarrow \infty$, then $R\rightarrow \frac{(\alpha+\beta)}{2C\lambda}$. Also for $\beta\neq\frac{8\pi}{3}$, $R$ vanishes for $ t=\frac{1}{2\lambda}ln\left(-\frac{\tau}{\sigma}\right)$.

\section{Summary and Conclusion}

We have presented two cosmological models of the Universe based on different forms of de-Sitter dynamics. The dynamical parameters for both the models are obtained and have been tested with the energy conditions criteria and the geometrical diagnostics. The wormhole throat radius of both the models are also calculated. Some of the key results are:
\begin{itemize}
\item The QDS and LCE scale factors are employed to analyse the dynamical behaviour of the EoS parameter. The effect of the parameters of both the QDS and LCE models are observed on the evolutionary aspects of the EoS parameter. While the LCE model reduces to the $\Lambda$CDM behaviour at the present epoch, the QDS model shows a departure from the  $\Lambda$CDM behaviour. The LCE model behaves like a Phantom field dominated model where the EoS parameter increases from a higher negative values to converge to $\omega=-1$ at late times. However, for the QDS model, for a substantial time zone, the EoS parameter appears to be a constant quantity close to $-1$ and then increases at late epoch. The effect of the coupling constant is quite visible in the dynamical behaviour of the EoS parameter. For the QDS model, with an increase in $\beta$, the EoS parameter increases but for the LCE model, $\omega$ at a given epoch decreases with $\beta$.  The interesting aspect of both the models are that, both the model predict the EoS parameter at the present epoch which are quite compatible to recent results.
\item In order to obtain the good dynamical and geometrical description of the cosmological model, the cosmographic coefficients are more useful. It involves the higher order derivative of the scale factor. The first two derivative at present epoch provides the Hubble parameter and deceleration parameter respectively. The third and fourth derivatives lead to the geometrical parameters $j$ and $s$.  The $(j,s)$ diagnostic pair for the LCE model approach to $(1,0)$ at late times indicating its compatiblity with the concordant $\Lambda$CDM behaviour. Also, the values predicted at the present epoch from LCE model are closer to  $\Lambda$CDM values $\{1,0\}$. However, for the QDS model, from the analysis of the statefinder diagnostics, we obtain a slight departure from the $\Lambda$CDM behaviour.

\item Energy conditions has a major role in identifying the physical frame \cite{Magnano94} of any theoretical model. So the physical meaning of the extended gravity depends on the energy conditions and initial conditions. In addition to the observations,  this might be a formidable tool to test the dark components of the Universe. In this work,  we have obtained and analysed the graphical behaviour of respective energy conditions of both the models that show a clear violation of SEC and satisfaction of the DEC. However, while the QDS model satisfies NEC, the LCE scenario violates the null energy conditions. This violation of NEC in LCE model leads to a phantom field like behaviour of the model. The result is as expected in the geometrically extended gravity \cite{Capozziello14, Capozziello19}.

\end{itemize}

As a final remark, we say that, the de Sitter like models as discussed in the present work may be useful in understanding the hidden mystery of the Universe concerning the late time cosmic acceleration issue as well as a possible bouncing feature at the initial epoch. Although, the QDS model shows a little departure from the $\Lambda$CDM behaviour, the LCE model seems to be compatible to the  $\Lambda$CDM model at least at the present epoch and is quite incompatible with $\Lambda$CDM behaviour in the recent past upto $z\simeq 3$.

\section*{Acknowledgements} BM and SKT acknowledge IUCAA, Pune (India) for providing the necessary support under the visiting associateship program. BM and EG acknowledge DST, New Delhi, India for providing facilities through DST-FIST lab, Department of Mathematics, where a part of this work was done. The authors are thankful to the anonymous referee for the constructive suggestions and comments for the improvement of the paper.


\section*{References}
\begin{thebibliography}{99} 

\bibitem{Perlmutter98} S. Perlmutter, et al., \textit{Nature}, \textbf{391}, 51 (1998).

\bibitem{Reiss98} A.G.Reiss et al., \textit{Astron. J.}, \textbf{116}, 1009 (1998).

\bibitem{Knop03} R. Knop et al., \textit{Astrophys. J.}, \textbf{598}, 102 (2003).

\bibitem{Eisenstein05} D.J. Eisenstein et al., \textit{Astrophys. J.}, \textbf{633}, 560 (2005).

\bibitem{Spergel07} D.N. Spergel et al., \textit{Astrophys. J.} , \textbf{170}, 377 (2007).

\bibitem{Sullivan11} M. Sullivan et al., \textit{Astrophys. J.}, \textbf{737}, 102 (2011).

\bibitem{Suzuki12} N. Suzuki et al., \textit{Astrophys. J.}, \textbf{746}, 85 (2012).

\bibitem{Nojiri06a} S. Nojiri, S. D. Odintsov, \textit{Phys. Rev. D}, \textbf{74}, 086005 (2006).

\bibitem{Capozziello11} S. Capozziello, M. De Laurentis, \textit{Phys. Reports}, \textbf{509}, 167 (2011).

\bibitem{Nojiri17} S.Nojiri, S.D.Odintsov, V.K.Oikonomou, \textit{Phys. Reports}, \textbf{692}, 1 (2017).  

\bibitem{Harko11} T. Harko et al., \textit{Phys. Rev. D}, \textbf{84}, 024020 (2011).

\bibitem{Nojiri06b} S. Nojiri, S. D. Odintsov, M. Sami, \textit{Phys. Rev. D}, \textbf{74}, 046004 (2006), D. Bazeia et al., \textit{Phys. Lett. B}, \textbf{649}, 445 (2007).

\bibitem{Obukhov03} Y. N. Obukhov, J. G. Pereira, \textit{Phys. Rev. D}, \textbf{67}, 044016 (2003), E.V. Linder, \textit{Phys. Rev. D.}, \textbf{81}, 127301 (2010).

\bibitem{Xu19} Y. Xu, G.Li, T. Harko, \textit{Eur. Phys. J. C}, \textbf{79}, 708 (2019).

\bibitem{Odintsov2020}S. D. Odintsov,  V. K. Oikonomou, \textit{Phys. Rev. D}, \textbf{101}, 044009 (2020).

\bibitem{Odintsov2020a}S. D. Odintsov, V. K. Oikonomou, F. P. Fronimos, K. V. Fasoulakos, \textit{Phys. Rev. D}, \textbf{102}, 104042 (2020).

\bibitem{Odintsov2021}S. D. Odintsov, V. K. Oikonomou, F. P. Fronimos, arXiv:2102.02239.

\bibitem{Oikonomou2021} V. K. Oikonomou, \textit{Phys. Rev. D}, \textbf{103}, 044036 (2021).

\bibitem{Oikonomou2014}V. K. Oikonomou, N. Karagiannakis, M. Park, \textit{Phys. Rev. D}, \textbf{91}, 064029 (2014).

\bibitem{Houndjo12} M.J.S. Houndjo, \textit{Int. J. Mod. Phys. D}, \textbf{21}, 1250003 (2012).

\bibitem{Azizi13} T. Azizi, \textit{Int. J. Theo. Phys.}, \textbf{52}, 3486 (2013).

\bibitem{Santos13} A.F. Santos, \textit{Mod. Phys. Lett. A}, \textbf{28}, 1350141 (2013).

\bibitem{Alvarenga13} F.G. Alvarenga, \textit{Phys. Rev. D}, \textbf{87}, 103526 (2013).

\bibitem{Pasqua13}  A. Pasqua, S. Chattopadhyay, I. Khomenko, \textit{Can. J. Phys.}, \textbf{91}, 632 (2013). 

\bibitem{Yadav14} A.K. Yadav, \textit{Eur. Phys. J. Plus}, \textbf{129}, 194 (2014).

\bibitem{Das16} A. Das et al., \textit{Eur. Phys. J. C}, \textbf{76}, 654 (2016).

\bibitem{Yousaf16} Z. Yousaf, K. Bamba, M. Z. H. Bhatti, \textit{Phys. Rev. D}, \textbf{93}, 124048 (2016).

\bibitem{Tripathy2019} S. K. Tripathy, R. K. Khuntia, P. Parida, \textit{Eur. Phys. J. Plus}, \textbf{134}, 504 (2019).

\bibitem{Zaregonbadi16} R. Zaregonbadi, M. Farhoudi, N. Riazi, \textit{Phys. Rev. D}, \textbf{94}, 084052 (2016).

\bibitem{Alhamzawi16} A. Alhamzawi, R. Alhamzawi, \textit{Int. J. Mod. Phys. D}, \textbf{25}, 1650020 (2016).

\bibitem{Fayaz16} V. Fayaz et al., \textit{Eur. Phys. J. Plus}, \textbf{131}, 22 (2016).

\bibitem{Xu16} M. Xu, T. Harko, S. Liang, \textit{Eur. Phys. J. C}, \textbf{76}, 449 (2016). 

\bibitem{Shabani17} H. Shabani, A.H. Ziaie, \textit{Eur. Phys. J. C}, \textbf{77}, 507 (2017).

\bibitem{Sharif17} M. Sharif, I. Nawazish, \textit{Eur. Phys. J. C}, \textbf{77}, 198 (2017).
 
\bibitem{Islam18} S. Islam, S. Basu, \textit{Chin. Phys. Lett.}, \textbf{35}, 099501 (2018).

\bibitem{Elizalde18} E. Elizalde, M. Khurshudyan, \textit{Phys. Rev. D}, \textbf{98}, 123525 (2018).

\bibitem{Mishra18} B. Mishra, S. Tarai, S.K. Tripathy, \textit{Mod. Phys. Lett. A}, \textbf{33}, 1850170, 2018.

\bibitem{Khan18} S. Khan, M.S. Khan, A. Ali, \textit{Mod. Phys. Lett. A}, \textbf{33}, 1850065 (2018).

\bibitem{Wu18} J. Wu et al., \textit{Eur. Phys. J. C}, \textbf{78}, 430 (2018). 

\bibitem{Barrientos18}E. Barrientos, \textit{Phys. Rev. D}, \textbf{97}, 104041 (2018).

\bibitem{Tretyakov19} P. V. Tretyakov, \textit{Eur. Phys. J. C}, \textbf{78}, 896 (2018).

\bibitem{Mishra19} B.Mishra, G. Ribeiro, P.H.R.S. Moraes, \textit{Mod. Phys. Lett A}, \textbf{34}, 1950321 (2019).

\bibitem{Fisher19} S. B. Fisher, E. D. Carlson, \textit{Phys. Rev. D}, \textbf{100}, 064059 (2019).

\bibitem{Tripathy20} S.K. Tripathy, B.Mishra, \textit{Chin. J. Phys.} \textbf{60}, 448 (2020). 

\bibitem{Maurya20} S.K. Maurya, F. Tello-Ortiz, \textit{Phys, Dark Universe}, \textbf{27}, 100442 (2020).

\bibitem{Mishra20a} B.Mishra, S.K. Tripathy, S. Ray, \textit{Int. J. Mod. Phys. D.}, \textbf{29}, 2050100 (2020).

\bibitem{Tripathy2020} S. K. Tripathy, \textit{Phys. of Dark Univ.}, \textbf{31}, 100757 (2021).

\bibitem{Rahaman20} M. Rahaman, \textit{Eur. Phys. J. C}, \textbf{80}, 272 (2020). 

\bibitem{Ahmad20} R. Ahmad, G. Abbas, \textit{Mod. Phys. Lett. A}, \textbf{35}, 2050103 (2020).

\bibitem{Mishra20b} B.Mishra, S.K. Tripathy, \textit{Phys. Scr.}, \textbf{95}, 095004 (2020).

\bibitem{Gamonal21} M. Gamonal, \textit{Phys. Dark Universe}, \textbf{31}, 100768 (2021).

\bibitem{Hawking73} S.W. Hawking, G.F.R. Ellis, \textit{The Large Scale Structure of Space–Time}, Cambridge Univ. Press, Cambridge (1973).

\bibitem{Tahim07} M.O. Tahim, R.R. Landim, C.A.S. Almeida, arXiv:0705.4120[gr-qc].

\bibitem{Visser97} M. Visser, \textit{Science}, \textbf{276}, 88 (1997).

\bibitem{Atazadeh09} K. Atazadeh et al., \textit{Int. J. Mod. Phys. D}, \textbf{18}, 1101 (2009).

\bibitem{Wang12} J. Wang, K. Liao, \textit{Class. Quantum Gravity}, \textbf{29}, 215016 (2012).

\bibitem{Banijmali12} A. Banijamali, B. Fazlpour, M.R. Setare, \textit{
Astrophys. Space Sci.}, \textbf{338}, 327 (2012). 

\bibitem{Esmaeili18} F. M. Esmaeili, B. Mishra, \textit{J. Astrophys. Astr.}, \textbf{39}, 59 (2018). 

\bibitem{Bhatti18} M. Z. Bhatti, Z. Yousaf, M. Ilyas, \textit{J. Astrophys. Astr.}, \textbf{39}, 69 (2018).

\bibitem{Capozziello18} S. Capozziello, S. Nojiri, S.D.Odintsov, \textit{Phys. Lett. B}, \textbf{781}, 99 (2018).

\bibitem{Camarena2020} D. Camarena, V. Marra, \textit{Phys. Rev. Res.}, \textbf{2}, 013028 (2020).

\bibitem{Goswami2021} G. K. Goswami, A. K. Yadav, B. Mishra, S. K. Tripathy, \textit{Fortschr. Phys.}, \textbf{2021}, 2100007 (2021).

\bibitem{Amanullah10} R. Amanullah et al., \textit{Astrophys. J.}, \textbf{716}, 712 (2010).

\bibitem{Hinshaw13} G. Hinshaw et al., \textit{Astrophys. J. Suppl. Ser.}, \textbf{208}, 19 (2013).

\bibitem{Kumar14} S. Kumar, L. Xu, \textit{Phys. Lett. B}, \textbf{737}, 244 (2014).

\bibitem{Aghanim20} N. Aghanim et al., \textit{Astronomy Astrophys.}, \textbf{641}, A6 (2020).

\bibitem{Babichev04} E.Babichev, V.Dokuchaev, Y. Eroshenko, \textit{Phys. Rev. Lett.}, \textbf{93}, 021102 (2004).

\bibitem{Astashenok12} A.V. Astashenok et al., \textit{Phys. Lett. B}, \textbf{709}, 396 (2012).

\bibitem{Tripathy2021} S. K. Tripathy, B. Mishra, S. Ray, R. Sengupta, \textit{Chin. J. Phys.}, \textbf{71}, 610 (2021).

\bibitem{Bamba2014} K. Bamba, A. N. Makarenko, A. N. Myagky, S. D. Odintsov, \textit{Phys. Lett. B}, \textbf{732}, 349 (2014).

\bibitem{Magnano94} G. Magnano, L.M. Sokolowski, \textit{Phys. Rev. D}, \textbf{50}, 5039 (1994).

\bibitem{Capozziello14} S. Capozziello, F.S.N.Lobo, J.P.Mimoso, \textit{Phys. Lett. B}, \textbf{730}, 280 (2014).

\bibitem{Capozziello19}S. Capozziello, R D'Agostino, O. Luongo, \textit{Int. J. Mod. Phys. D}, \textbf{28}, 19350016 (2019).






\end{thebibliography}
\end{document}